\documentclass{iau307}
\usepackage{graphicx}
\usepackage{natbib}
\bibpunct{(}{)}{;}{a}{}{,}

\title[UVMag] %% give here short title %%
{The UVMag space project: UV and visible spectropolarimetry of massive stars
}

\author[C. Neiner and the UVMag consortium]   %% give here short author list %%
{Coralie Neiner
\and
the UVMag consortium}

\affiliation{LESIA, Observatoire de Paris, CNRS UMR 8109, UPMC, Universit\'e
Paris Diderot,\\
5 place Jules Janssen, 92190 Meudon, France\\
email: {\tt coralie.neiner@obspm.fr}}

\pubyear{2014}
\volume{307} 
\pagerange{}
\jname{New windows on massive stars: asteroseismology, interferometry, and spectropolarimetry}
\editors{G. Meynet, C. Georgy, J.H. Groh \& Ph. Stee, eds.}
\begin{document}

\maketitle

\begin{abstract}
UVMag is a medium-size space telescope equipped with a high-resolution
spectropolarimetrer working in the UV and visible domains. It will be proposed
to ESA for a future M mission. It will allow scientists to study all types of
stars as well as e.g. exoplanets and the interstellar medium. It will be
particularly useful for massive stars, since their spectral energy distribution
peaks in the UV. UVMag will allow us to study massive stars and their
circumstellar environment (in particular the stellar wind) spectroscopically in
great details. Moreover, with UVMag's polarimetric capabilities we will be able,
for the first time, to measure the magnetic field of massive stars
simultaneously at the stellar surface and in the wind lines, i.e. to completely
map their magnetosphere.
\end{abstract}

\firstsection % if your document starts with a section,
\section{The UVMag space mission}

UVMag is a space mission dedicated to the study of the dynamic 3D environment of
stars and planets. It will consist in a 1.3-meter telescope and  a
spectropolarimeter covering the UV and visible wavelength range from 117 to 870
nm. An option for far-UV (90-117 nm) is also currently explored. The spectral
resolution will be at least 25000 in the UV domain and at least 35000 in the
visible domain. Full Stokes (IQUV) information, i.e. both circular and linear
polarisation, will be obtained. More details on the spectropolarimeter design 
are available in \cite{pertenais2014}. The mission will be proposed to ESA this
autumn as its M4 mission, for a launch in 2025. It will last 5 years.

\section{What can UVMag do for massive stars?}

UVMag is particularly well suited for massive stars since it will observe in the
UV and visible domains, i.e. where massive stars emit most of their lines. UVMag
will mainly target massive stars with magnitudes between V=3 and 10, but massive
stars in the Magellanic Clouds will also be reachable with a longer exposure
time. This will allow us to probe stars in a different environment.  UVMag will
allow us to:
\begin{itemize}
\item study the stellar wind through UV resonance lines and visible
recombination lines, in particular for O stars;
\item study wind clumping and the line-driven instabilities;
\item study the magnetic field at the stellar surface with improved
signal-to-noise. Indeed, thanks to the increased number of photons and of lines
available in the UV compared to the visible domain, the signal-to-noise ratio of
Zeeman magnetic signatures obtained with the LSD technique in the UV is higher
than in the visible domain; 
\item study the magnetosphere and confinement of material around the star.
Emission from the plasma trapped in the magnetosphere can be observed in visible
emission lines, while the confined wind can be studied in the UV resonance
lines; 
\item study linear polarisation and depolarisation effects from circumstellar
disks.
\end{itemize}
With UVMag we will thus be able to study the formation, evolution and
environment of massive stars. More details about UVMag's science case for
massive stars as well as for other topics can be found in \cite{neiner2014}.

\section{UVMag's observing program}

UVMag will observe three types of targets:
\begin{itemize}
\item Mapping targets: 50 to 100 stars (of all types) will be followed over at
least one full rotation period with high cadence in order to study them in great
details and reconstruct 3D maps of their surface and environment. These targets
will be partly secured through the consortium core program and partly chosen
following a competitive proposal process. Some targets (in particular solar-type
stars) will be reobserve every year to study their variability over activity
cycles.
\item Survey targets: several thousands stars will be observed once or twice to
provide information on their magnetic field, wind and environment. This will
include an unbiased magnitude-limited statistical sample and targets selected
through a competitive proposal process. These snapshot data will provide
statistical results as well as specific inputs (e.g. wind terminal velocity) for
stellar modelling.
\item A Target of Opportunity (ToO) mode is also planned, in particular for
supernovae and outbursting stars such as classical Be stars.\\
\end{itemize}

There are $\sim$50000 stars with 3$<$V$<$10 observable with UVMag. Among them,
there are $\sim$20000 OB stars. Since 7\% of OB stars are found to be magnetic
\citep{wade2013}, there are statistically $\sim$1400 magnetic OB stars among the
$\sim$20000 OB stars. However, only $\sim$100 magnetic OB stars are known as of
today. Although this number has been growing significantly since the new
generation of spectropolarimeters (Narval at TBL, ESPaDOnS at CFHT and HarpsPol
at ESO) is available, it will probably remain rather low in the coming decade.
It is therefore probable that all magnetic OB stars known at the time of launch
will be observed by UVMag in the survey sample (and several of them will be
mapped in details). Of course, non-magnetic OB stars will also be observed.

For massive stars, the signal-to-noise ratio in the intensity spectrum will be
above 100 in 20 minutes exposure, both in the UV and visible domains. 

\section{Conclusions}

UVMag is an M-size space mission with a 1.3 meter telescope equipped with a
high-resolution spectropolarimeter working in the UV and visible domain
simultaneously. This mission will be particularly useful for the study of
massive stars which emit most of their light in this wavelength domain. This
includes the study of their wind, magnetic field, magnetosphere, disk, clouds as
well as of their surface (e.g. spots).

\bibliographystyle{iau307}
\bibliography{IAUS307Neiner2}

\end{document}